\documentclass[12pt]{article}

\usepackage{amsfonts}
 \usepackage{amssymb}
 \usepackage{amsmath}

\def\pmx{\begin{pmatrix}}
\def\emx{\end{pmatrix}}
\def\bsq{\begin{subequations}}
\def\esq{\end{subequations}}

\def\be{\begin{eqnarray}}
\def\ee{\end{eqnarray}}
\def\bee{\begin{eqnarray*}}
\def\eee{\end{eqnarray*}}

\newtheorem{thm}{Theorem}

\newtheorem{defn}[thm]{Definition}

 \def\tr{{\rm {Tr}} \, }

\def\bra{\langle}
\def\ket{\rangle}

\def\ot{\otimes}

\def\b0{{\mathbf{\bold 0}}}

\title{Correcting quantum channels by measuring the environment}

 \author{Patrick Hayden
 \\ Institute for Quantum Information
 \\ Caltech 107-81
 \\ Pasadena CA 91125
 \\ and
 \\ Christopher King
\\ Department of Mathematics
\\ Northeastern University
\\ Boston MA 02115
\\
}

\begin{document}

\maketitle

\begin{abstract}
The corrected capacity of a quantum channel is defined as the best one-shot
capacity that can be obtained by measuring the environment and using the
result to correct the output of the channel. It is shown that (i) all qubit channels
have corrected capacity $\log 2$, (ii) a product of $N$ qubit  channels has corrected
capacity $N \log 2$, and (iii) all channels have corrected capacity at least $\log 2$.
The question is posed of finding the channel with smallest corrected capacity in any dimension $d$.
\end{abstract}

\section{Introduction and statement of results}
Every quantum channel can be viewed as arising from the unitary interaction of a system
with its environment. The resulting entanglement between system and environment is lost
 when the environment is `traced out', thereby destroying the  purity of the signal states
 and introducing noise into the system. Specifically, letting $\cal S$ denote the
system and $\cal E$ its environment, the action of the channel $\Phi$ on $\cal S$ is obtained as
\be\label{def:Phi}
\Phi (\rho) = {\tr}_{\cal E} \, \bigg[ U_{\cal S E} \,( \rho \ot \omega ) \, U_{\cal S E}^{*} \bigg]
\ee
where the unitary matrix $U_{\cal S E}$ entangles the system and environment, and $\omega$
is a state in $\cal E$.

In a recent paper, Gregoratti and Werner \cite{GW} explored the
extent to which the noise produced by the channel $\Phi$ could be
removed by performing a measurement on the environment and using
the result to correct the output state of the channel.  To be
specific, let $\{ X_k \}$ denote a POVM acting on the environment.
If the result `$k$' is obtained from this measurement, then the
output state is, up to normalization,
 \be
 {\tr}_{\cal E} \, \bigg[ (I \ot X_{k}) \,U_{\cal S E} \,( \rho \ot \omega )\, U_{\cal S E}^{*} \bigg] = A_k \, \rho \, A_{k}^{*}
 \ee
where this expression defines the matrix $A_k$. If the measurement result is ignored, then the output state is just
 \be
 \sum_{k} \, A_k \, \rho \, A_{k}^{*} = \Phi(\rho).
 \ee
 In fact every Kraus representation of $\Phi$ arises in this way, as the effect of an unrecorded measurement on the environment. However, if the result of the measurement {\em is}
 recorded, then there is the possibility of making a correction to the output state, based on the measurement result. That is, one could apply a completely positive trace preserving map $R_k$ to the output, conditioned on receiving the result $k$ from the measurement.
The resulting output state would then be
 \be
 \sum_{k} \, R_k ( A_k \rho A_{k}^{*} ).
 \ee
  Writing ${\cal A} = \{A_1, \dots, A_N\}$ and
 ${\cal R} = \{ R_1, \dots , R_N \}$, this defines a new channel, which is  a corrected
 version of $\Phi$, namely
 \be
  {\Phi}_{\cal A,R}(\cdot) = \sum_{k} \, R_k ( A_k \, \cdot \, A_{k}^{*} ).
\ee

 This corrected channel ${\Phi}_{\cal A,R}$ may be less noisy than $\Phi$ if the maps
 $R_k$ are chosen well. For example, if $\Phi$ has a Kraus representation with operators
 $A_k = \sqrt{p_k} \, V_k$, where the $\{ V_k \}$ are unitary and $\sum p_k = 1$,
 then by choosing $R_k (\cdot) = V_{k}^{*} (\cdot) V_k$ the channel can be corrected to the identity, that
 is ${\Phi}_{\cal A,R} = I$ in this case. This is an extreme case of course, and in fact
 Gregoratti and Werner show that this can happen if and only if $\Phi$ is such a `random
 unitary' channel.

 Nevertheless this example raises the question of determining the `best' correction that
 can be achieved for a given channel $\Phi$. We will use the $1$-shot Shannon capacity of the
 corrected channel as a way to quantify `best'. That is, we consider the optimal
 combination of input states and output measurements for the
channel $\Phi_{\cal A,R}$, in order to maximize the mutual information between input
and output. This maximum mutual information is the Shannon capacity of the
corrected channel $C_{\rm Shan}(\Phi_{{\cal A,R}})$. Furthermore,
in order to find the overall best correction for $\Phi$, we must
maximize over choices of $\cal R$ to find the best correction for any Kraus
representation of $\Phi$, and then maximize this quantity
 over the choice of Kraus operators. Accordingly, we denote by
${\cal K}(\Phi)$ the collection of all Kraus sets for $\Phi$, that
is all collections  $\{A_1, \dots, A_N\}$ satisfying
\be
\sum_{k=1}^N  A_{k}^{*} \, A_k = I
\ee
and
\be
\Phi(\rho) = \sum_{k=1}^N A_k \, \rho \, A_{k}^{*}.
\ee
Notice that different elements of
${\cal K}(\Phi)$ may contain different numbers of matrices.

\begin{defn}
The optimal corrected capacity for $\Phi$ is
\be
C_{\rm corr}(\Phi) = \sup_{{\cal A} \in {\cal K}(\Phi)} \,\sup_{\cal R} \, C_{\rm Shan}(\Phi_{{\cal A,R}}) .
\ee
\end{defn}

\medskip
It is nearly immediate, for example, that the optimal corrected
capacity of a so-called classical-quantum (c-q) channel~\cite{H}
on ${\bf C}^d$ is $\log d$. By definition, a c-q channel $\Phi$
can always be written in the form
\begin{equation}
\Phi(\rho) = \sum_k \langle k | \rho | k \rangle \, \sigma_k
\end{equation}
for a set of density operators $\{ \sigma_k \}$ and orthonormal
basis $\{ | k  \rangle \}$. One possible choice for the operators
$R_k$ is then to set each to the constant map $R_k(\sigma) = | k
\rangle\langle k |$, in which case
\begin{equation}
\Phi_{{\cal A}, {\cal R}}(\rho) = \sum_k | k \rangle\langle k |
\rho | k \rangle\langle k |,
\end{equation}
which obviously has Shannon capacity $\log d$.

\medskip
We can now state our first result.

\begin{thm}\label{thm1}
For any qubit channel $\Phi$,
\be\label{thm1:eqn}
C_{\rm corr}(\Phi) = \log 2
\ee
\end{thm}

\medskip
As (\ref{thm1:eqn}) shows, every qubit channel can be corrected to a channel
with full capacity
by performing a measurement on the environment and correcting the output
depending on the result. The fact that this  is true for the completely noisy channel
$\Phi(\rho) = 1/2 \,\,I$ for example may seem surprising -- however it reflects the fact that
the information about the initial state is stored in the environment and in this case can be fully recovered
by measurement.

\medskip
Theorem \ref{thm1} is proved by showing that for any qubit channel $\Phi$ it is possible to find
two orthogonal input states which can be perfectly distinguished by making measurement-based
corrections at the output. In fact this result holds in any dimension, and therefore provides
the same lower bound on the optimal corrected capacity for any channel.

\medskip
\begin{thm}\label{thm2}
For any channel $\Phi$,
\be\label{thm2:eqn}
C_{\rm corr}(\Phi) \geq \log 2
\ee
\end{thm}

\medskip
\par\noindent{\em Remarks}
\par\noindent {\bf 1)}
One could  consider other measures of `best' correction for a channel, for example
the Holevo capacity. However, operationally this refers to making entangled measurements
on outputs from multiple copies of the channel, and in this case it probably makes sense
to also consider corrections which arise from entangled measurements on multiple
copies of the environment, so this should be done in a more general setting.

\medskip
\par\noindent {\bf 2)} For a qubit channel Theorem \ref{thm1} says that it is always possible to achieve
full transmission capacity by measuring the environment and applying corrections to the
channel output. It follows that the same is true for a product of qubit channels, and furthermore
this can be done by
making independent measurements on the environment of each qubit.

\medskip
\par\noindent {\bf 3)}
In dimensions higher than two, the bound in (\ref{thm2:eqn}) is certainly not
tight. However it remains an open question to find a larger bound.
For each dimension $d$ there is a worst-case
channel (or channels) for which $C_{\rm corr}(\Phi)$ takes its smallest value, so we could define
\be
C_{\rm corr}(d) = \inf \, \{ C_{\rm corr}(\Phi) \,:\, \Phi \, \mbox{is CPT on} \, {\bf C}^d \} .
\ee
Then the question becomes:
what are these channels, and what are these worst values?

\section{Proof of Theorems}
Theorem \ref{thm1} is a special case of Theorem \ref{thm2}, and Theorem \ref{thm2} can be deduced
from the following  result of Walgate et al \cite{Wa}:
any pair of orthogonal pure states in  a bipartite system can be perfectly distinguished
using LOCC. So if we use two orthogonal signal states
$| \psi_1 \ket$ and $| \psi_2 \ket$ for the channel, then
the entangled states $U_{\cal S E} | \psi_1 \ket$ and $U_{\cal S E} | \psi_2 \ket$ are orthogonal
and hence can be perfectly distinguished by first measuring in $\cal E$, then using the result to
select a measurement in $\cal S$. Hence the capacity of this corrected channel
is at least $\log 2$, and this proves the Theorem.

For completeness we include  a direct proof of Theorem \ref{thm2}.
The key idea is to find  a Kraus representation
$A_1, \dots, A_N$ for $\Phi$ with the property that the first and second columns of every matrix $A_k$  are orthogonal, and to use the first two canonical basis vectors
$| e_1 \ket$ and $| e_2 \ket $ as the signal states. Measuring the value `$k$' on these states will produce either $A_k  | e_1 \ket \bra e_1 | A_{k}^{*}$ or
$A_k  | e_2 \ket \bra e_2 | A_{k}^{*}$, and these are the projections onto the
first and second column vectors of $A_k$ respectively. By assumption these are orthogonal, and therefore can be perfectly distinguished.

So the proof reduces to showing that every channel has a Kraus representation
with this property. To show this, let $A_1, \dots, A_N$ be any Kraus representation
for $\Phi$, and define the $N \times N$ matrix $M(A)$ by
\be
M(A)_{ij} = \tr A_i \, | e_{1} \ket \bra e_{2} | \, A_{j}^{*} =
\bra e_{2} | \, A_{j}^{*} \, A_i \, | e_{1} \ket
\ee
So $M(A)_{ij}$ is the inner product of the first column of $A_i$ with the second
column of $A_j$. Now let $V$ be any unitary
$N \times N$ matrix, and define the matrices
\be
B_i = \sum_{j=1}^N V_{ij} \, A_j .
\ee
Then $B_1, \dots, B_N$ is also a Kraus representation for $\Phi$. Furthermore
\be
M(B) = V M(A) V^{*} .
\ee
We now use the following interesting mathematical fact \cite{HJ}: given the matrix $M(A)$, there
is a unitary matrix $V$ so that all diagonal entries of $M(B)$ are equal. Since
$\sum A_{i}^{*} A_i = I$ it follows that $\tr M(A) = 0$. Hence with this choice of $V$,
all diagonal entries of $M(B)$ are zero. This means that for every matrix $B_i$, the
first and second columns are orthogonal, and so $B_1, \dots, B_N$
is the desired representation.

\bigskip
{\bf Acknowledgements} PH is supported by the Sherman Fairchild
Foundation as well as the National Science Foundation through
grant EIA-0086038. CK thanks the Institute for Quantum Information
for their hospitality while this work was completed. CK's research
was supported in part by National Science Foundation Grant
DMS-0400426.

{~~}


\bigskip

\begin{thebibliography}{~~}


\bibitem{GW}  M. Gregoratti  and R.~F. Werner,
``On quantum error correction by classical feedback in discrete time'',
quant-ph/0403092.

\bibitem{H} A. Holevo, ``Coding Theorems for Quantum Channels'',
{\em Russian Math Surveys} {\bf 53}, 1295-1331 (1999).

\bibitem{HJ} R. Horn and C. Johnson,
``Matrix Analysis" Section 2.2, Exercise 3,
Cambridge University Press, 1985.

\bibitem{Wa} J. Walgate, A.~J. Short, L. Hardy, V. Vedral,
``Local Distinguishability of Multipartite Orthogonal Quantum States'',
{\em  Phys. Rev. Lett.} {\bf 85},  4972 (2000).


\end{thebibliography}
\end{document}